\documentclass[twocolumn,preprintnumbers,letterpaper,nofootinbib,amsmath,amssymb]{revtex4}

\usepackage{mathtools}
\usepackage{graphicx,xcolor,tikz}
\usetikzlibrary{decorations.pathmorphing}
\tikzset{snake it/.style={decorate, decoration=snake}}

\DeclareMathOperator{\re}{Re}
\DeclareMathOperator{\im}{Im}

\newcommand{\revone}[1]{#1}

\newcommand{\revtwo}[1]{#1}

\newcommand{\revthr}[1]{#1}

\usepackage{hyperref}
\usepackage{orcidlink}

\begin{document}

\preprint{RIKEN-iTHEMS-Report-25}

\title{On continuum and resonant spectra from exact WKB analysis}

\author{Okuto Morikawa \orcidlink{0000-0002-0044-4491}}
\email{okuto.morikawa@riken.jp}
\affiliation{Center for Interdisciplinary Theoretical and Mathematical Sciences (iTHEMS), RIKEN, Wako 351-0198, Japan}

\author{Shoya Ogawa \orcidlink{0000-0003-0900-2486}}
\email{ogawa.shoya.615@m.kyushu-u.ac.jp}
\affiliation{Department of Physics, Kyushu University, 744 Motooka, Nishi-ku, Fukuoka 819-0395, Japan}

\begin{abstract}
Resonance phenomena are central to many quantum systems, where resonant states are typically characterized by pole singularities of the S-matrix. In this work, we employ the complex scaling method (CSM) in conjunction with exact WKB analysis to elucidate the geometric structure of scattering problems that encompass both bound and resonant states. By analyzing the continuum spectrum via the exact WKB framework, we derive the S-matrix for the inverted Rosen--Morse potential and reveal its underlying complex-geometric features. Furthermore, we reinterpret the Aguilar--Balslev--Combes theorem, the foundation of CSM, from a geometric perspective, and discuss the physical significance of the Siegert boundary condition within a rigorously defined modified Hilbert space. Our analysis bridges scattering cross-sections and spectral theory, offering new geometric insights into quantum resonance and scattering phenomena.
\end{abstract}

\maketitle

\tableofcontents

\section{Introduction}
Resonance phenomena manifest universally across physics, chemistry, and mathematics, typically corresponding to unstable or quasi-stationary states that transcend the scope of perturbation theory. Quantum resonance, in particular, occupies a central position in diverse modern physical contexts, from nuclear reactions and molecular scattering to optical cavities and condensed-matter excitations. A coherent theoretical characterization arises from the pole singularities of the S-matrix or the resolvent of the Hamiltonian, which signal the existence of resonant states in the complex energy plane. Nevertheless, any comprehensive and rigorously established formulation remains intrinsically hindered when viewed from alternative perspectives, including functional, scattering, probabilistic, and operator-algebraic frameworks.

\begin{figure}[t]
 \centering
  \begin{tikzpicture}[x=0.6mm,y=0.6mm,>=latex]
   \draw[->,line width=2pt] (-30,0) -- (30,0);
   \draw[->,line width=1pt] (0,-28) -- (0,28);
   \draw[line width=1pt] (-25,20) -- (-20,20) -- (-20,25);
   \draw[line width=1pt] (-22.5,22.5) node {$k$};
   \draw[line width=1pt,blue] (8,-5) circle (2);
   \draw[line width=1pt,blue] (16,-8) circle (2);
   \draw[line width=1pt,blue] (24,-13) circle (2);
   \fill (0,10) circle (2);
   \fill (0,18) circle (2);
   \draw[<-] (1.8,19.5) -- (10,22) node[right] {Bound};
   \draw[<-,blue] (16,-10.5) -- (15,-15) node[below] {Resonance};
   \node[below] at (0,-30) {(a) Complex $k$-plane};
  \end{tikzpicture}
\hspace{1em}
  \begin{tikzpicture}[x=0.6mm,y=0.6mm,>=latex]
   \draw[line width=1pt] (22,25) .. controls (22,0) and (30,0) .. (30,-25);
   \draw[line width=1pt] (30,28) .. controls (30,0) and (22,0) .. (22,-28);
   \draw[line width=1pt] (22,25) -- (-30,25) -- (-30,-28) -- (22,-28);
   \draw[line width=1pt] (30,28) -- (-22,28) -- (-22,25);
   \draw[line width=1pt] (30,-25) -- (22,-25);
   \draw[line width=1pt] (-27,0) -- (-4,0);
   \draw[->,line width=2pt] (-4,0) -- (26,0);
   \draw[->,line width=1pt] (-4,-24) -- (-4,24);
   \draw[line width=1pt] (-26,17) -- (-21,17) -- (-21,22);
   \draw[line width=1pt] (-23.5,19.5) node {$E$};
   \draw[blue,dashed,line width=1pt] (2,-5) circle (2);
   \draw[blue,dashed,line width=1pt] (10,-10) circle (2);
   \draw[blue,dashed,line width=1pt] (18,-20) circle (2);
   \fill (-13,0) circle (2);
   \fill (-23,0) circle (2);
   \draw[<-] (-23,2) -- (-22.5,5) node[above] {Bound};
   \draw[blue,<-] (9.5,-12) -- (8,-20) node[below] {Resonance};
   \node[below] at (0,-30) {(b) Complex $E$-plane};
  \end{tikzpicture}
 \caption{Distribution of S-matrix poles in (a) the complex $k$-plane and (b) the complex $E$-plane. In the right panel~(b), the bound states exist in the first Riemann surface, while the resonant states appear in the second Riemann surface.}
 \label{fig:pole}
\end{figure}
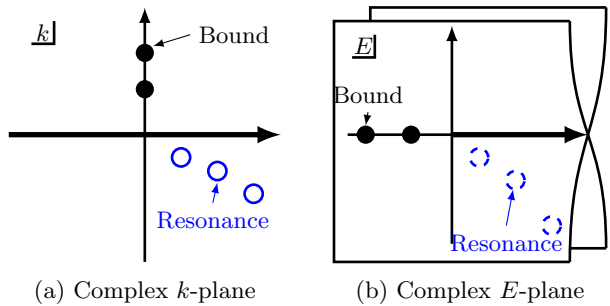

A resonant state has a complex wave number~$k$ or a complex energy~$E$, which causes the divergence of the asymptotic wave function, as shown in Fig.~\ref{fig:pole}. To rigorously describe resonant states, the complex scaling method (CSM) is often employed, which has proven successful in a variety of fields from atomic and molecular physics~\cite{Reinhardt:1982,Moiseyev:1979,Herbst:1981ew,Chu:1990,Telnov:2013}, to nuclear physics~\cite{Myo:2014ypa,Myo:2023heu,Guo:2010zzm,Matsumoto:2010mi,Kruppa:2013ala}, and hadron physics~\cite{Dote:2017,Song:2024ngu,Zhuo:2021,Lin:2024}.
Born in theoretical physics and refined through the insights of quantum chemistry, CSM has evolved into a powerful tool for the study of generic quantum resonances.

It is crucial in CSM to make resonant energies computable based on the so-called Aguilar--Balslev--Combes (ABC) theorem~\cite{Aguilar:1971ve,Balslev:1971vb}. This mathematical theorem states that the solutions of the Schr\"odinger equation possess the following properties under the complex scaling:
\begin{enumerate}
    \item The resonance solutions are described by the square-integrable functions, like bound states.
    \item The energies of the bound and resonant states are invariant with respect to the scaling.
    \item The continuum spectra start at the threshold energies of decays of the system into its subsystems and are rotated
    clockwise by an angle of 2$\theta$ from the positive real energy axis.
\end{enumerate}
The resonant state at a finite $\theta$ can be reduced to an eigenvalue problem, similar to a bound state. However, there are many subtleties because of its complex probability~\cite{Berggren:1996}, transition cross-section~\cite{Berggren:1971,Berggren:1978}, and no $*$-algebraic structure.

Meanwhile, the non-perturbative analysis of quasi-stationary states in quantum mechanics (QM) or quantum field theory (QFT) itself has long been recognized as an important issue. Resurgence theory regards the solution as an analytic function of a complexified perturbation parameter, and has been developed vigorously in recent years to give it a mathematically rigorous and physically meaningful sense.
See Refs.~\cite{Witten:2010zr,Aniceto:2018bis,Dorigoni:2014hea} for QM, and Refs.~\cite{Witten:2010cx,Dunne:2002rq,Dunne:2015eaa,Argyres:2012ka,Morikawa:2020agf,Bajnok:2021dri,Bajnok:2021zjm} for QFT.\footnote{See also famous findings in 1980s in Refs.~\cite{Bogomolny:1980ur,Zinn-Justin:1981qzi,David:1982qv}}
Particularly, the exact WKB analysis~\cite{Voros:1983xx,Delabaere:1999xx,Iwaki:2014vad}, which is a non-perturbative formulation of differential equations based on a formal higher-order WKB approximation and Borel resummation, provides a quite powerful toolkit for $1$-dimensional Schr\"odinger equations~\cite{Sueishi:2020rug,Sueishi:2021xti,Kamata:2021jrs,Kamata:2023opn,Kamata:2024tyb,Nikolaev:2008,Nikolaev:2021xzt,Nikolaev:2024,Kamata:2025dkk}.\footnote{There are many studies of differential equations in thermodynamic Bethe ansatz equation~\cite{Ito:2018eon,Ito:2019jio,Emery:2020qqu,Ito:2023cyz,Ito:2024nlt}, Painl\'eve equation~\cite{Basar:2015xna,Iwaki:2019zeq,Hollands:2019wbr,Imaizumi:2020fxf,vanSpaendonck:2022kit,DelMonte:2022kxh}, supersymmetry (e.g., Seiberg--Witten curve)~\cite{Kashani-Poor:2015pca,Kashani-Poor:2016edc,Ashok:2016yxz,Ashok:2019gee,Yan:2020kkb,Imaizumi:2021cxf,Grassi:2021wpw,Bianchi:2021xpr,Alim:2022oll,Imaizumi:2022qbi,Imaizumi:2022dgj}, and cosmology~\cite{Enomoto:2020xlf,Enomoto:2021hfv,Enomoto:2022mti,Enomoto:2022nuj,Honda:2024aro,Miyachi:2025ptm,Hatsuda:2019eoj,Hatsuda:2021gtn,Matyjasek:2019eeu,Decanini:2011eh,Alfaro:2024tdr,Motl:2003cd}. (See also Refs.~\cite{Taya:2020dco,Bucciotti:2023trp,Suzuki:2023slp}.)}

Despite the ubiquity of resonance, it has been profound and persistent whether the exact WKB analysis can be a mathematically precise and physically transparent framework. Recent works~\cite{Morikawa:2025grx,Morikawa:2025xjq} have demonstrated that the complex energy of resonance can be precisely quantized through exact WKB analysis, which has proven to be a powerful tool in studying quasi-stationary states. The resonant state is ``normalizable'' in exact WKB framework, and is incorporated in phenomenology, CSM, and spectral theory.

In this \revone{paper}, let us consider scattering problem and include continuum spectrum via the exact WKB analysis. Provided that the asymptotic behavior of wave functions is a plane wave as a boundary condition, we see the transmission and reflection waves according to appropriate analytical continuations and Stokes phenomena. Then, from the exact WKB analysis, we have the S-matrix explicitly in the case of the inverted Rosen--Morse potential. In general, any continuum state is ordinary normalized with the Jost function~\cite{Newton:1960}, and the S-matrix is also described by it. Our purpose is not to obtain new closed-form expressions for scattering coefficients, many of which are already available in the literature, but to provide a systematic and unified framework for understanding continuum and resonant spectra. Eventually, along with Refs.~\cite{Morikawa:2025grx,Morikawa:2025xjq}, every spectrum is collected on a Hilbert space with a self-adjoint Hamiltonian.\footnote{In this case, the corresponding physical states (bound and continuum spectra with real energies) in the Hilbert space do not include any resonance, but the Hamiltonian resolvent has pole singularities at resonances.}

Next, if introducing CSM, the physical meaning of all states becomes subtle but the exact WKB can define these states in a rigorous way. We propose a physical strategy to describe the ABC theorem. A particular wave function from the continuum spectrum may be divergent due to $\theta$ and $\arg E$, where $E$ corresponds to the complex energy of a resonant state. We then discuss the physical significance of the Siegert boundary condition, which prohibits incoming waves~\cite{Siegert:1939}, reflecting the irreversible nature of the decay process associated with resonances. Then, under the CSM, it is natural to give a basis of Hilbert space spanned by modified bound, resonant, and continuum states.

\section{Why Complex Scaling and Exact WKB Are Used}

\revone{In a textbook $1$-dimensional scattering problem with a potential barrier, the transmission and reflection coefficients can indeed be obtained by imposing continuity of the wavefunction (and its derivative) and using the probability-current continuity equation. If the goal is \emph{only} to compute $(R,T)$ at \textit{real} energies for such simple potentials, then neither complex scaling nor WKB is strictly necessary.}

\revone{However, the paper’s goal is broader: it is aimed at a unified treatment of \textit{continuum scattering} and \textit{resonances} (quasi-bound states), and at making the resonance problem mathematically well-posed and analytically tractable. That is precisely where complex scaling and (exact) WKB become essential.}

\subsection{\revone{Why complex scaling is used}}

\revone{Resonances correspond to poles of the S-matrix (or the resolvent) at \textit{complex} energies,
\begin{equation}
E = E_R - \frac{i}{2}\Gamma,
\end{equation}
with $\Gamma>0$ the decay width. We impose the ``purely outgoing'' boundary condition, called Siegert boundary condition \cite{Siegert:1939}.
The corresponding outgoing solutions typically behave asymptotically like
\begin{equation}
    \psi(x)\sim
    \begin{cases}
        e^{+ikx}& x\to +\infty,\\
        e^{-ikx}& x\to -\infty,
    \end{cases}
\end{equation}
but with $k=\sqrt{2mE}/\hbar$ \textit{complex}. This generally makes $\psi$ grow exponentially in at least one direction, so the resonance wavefunction is not square-integrable:
\begin{equation}
\psi\notin L^2(\mathbb{R}).
\end{equation}
Standard methods produce $R(E)$, $T(E)$ for real $E$, but they do not by themselves convert resonances into a discrete, well-defined eigenvalue problem.}

\revone{Complex scaling rotates the coordinate into the complex plane,
\begin{equation}
x\mapsto x e^{i\theta},\qquad 0<\theta<\frac{\pi}{4},
\end{equation}
which induces a non-unitary transformation $U_{\theta}$ (dilatation operator rather than translation operator) of the Hamiltonian $H$ to a complex-scaled operator $H_\theta=U_{\theta}HU_{\theta}^{-1}$. Under suitable conditions (captured by the ABC framework), the continuum spectrum is rotated in the complex energy plane, while resonance poles become \textit{isolated} eigenvalues of $H_\theta$ that are stable under changes of $\theta$. Operationally, this means:
\begin{itemize}
\item the outgoing, exponentially divergent resonance solutions become square-integrable after scaling;
\item resonances can be computed as discrete eigenpairs $(E,\psi_\theta)$ of $H_\theta$,
      rather than inferred indirectly from sharp peaks in $T(E)$ on the real axis.
\end{itemize}
So complex scaling is not mainly a computational shortcut; it is a way to make the resonance sector part of a spectral problem on a (modified) Hilbert-space setting.\footnote{\revone{Complex scaling should be understood as a mathematical analytic-dilation technique rather than a physical deformation of space. One performs a non-unitary similarity transformation corresponding to $x\mapsto x e^{i\theta}$, which defines a rotated operator $H_\theta$ when the potential is analytic (dilation-analytic) in a suitable sector. This rotation leaves the physical scattering problem on the real axis unchanged, but it renders purely outgoing resonant solutions square-integrable and exposes resonances as discrete eigenvalues of $H_\theta$ (while the continuous spectrum is rotated into the complex plane). In this sense, the ``complexification of coordinates'' provides a rigorous analytic continuation tool to access resonance poles of the S-matrix/resolvent, not a prescription for realizing complex coordinates in real space.
}}}

\subsection{\revone{Why exact WKB is used}}

\revone{For smooth potentials (such as the inverted Rosen--Morse potential), one can always solve the Schr\"odinger equation numerically, but this paper is concerned with analytic structure:
\begin{itemize}
\item how scattering data and resonance poles arise through analytic continuation;
\item how connection formulas change across turning points and Stokes lines.
\end{itemize}
Exact WKB (formal WKB plus Borel resummation), provides a systematic way to track the wavefunction globally in the complex plane and to derive the S-matrix structure from this analytic continuation.}

\revone{Resonance widths often scale like exponentially small tunneling factors (schematically),
\begin{equation}
\Gamma \sim e^{-2S/\hbar},
\end{equation}
where $S$ is a barrier action integral. Such effects are non-perturbative in $\hbar$ and can be subtle to extract robustly from purely real-axis data. Exact WKB is designed to control precisely these exponentially small contributions and their dependence on analytic continuation (Stokes jumps), which is closely tied to resonance formation.}

\subsection{\revone{Unified picture of scattering and spectra}}
\revone{A central motivation is that
a scattering problem includes \textit{resonances} (complex energies, outgoing boundary conditions, widths) and we want a unified, mathematically controlled treatment of continuum and resonant spectra, then complex scaling and exact WKB are not optional---they are the natural tools.
Standard methods efficiently compute $(R,T)$ on the real axis, but they do not naturally encode:
\begin{itemize}
\item the location of complex poles and their stability properties;
\item the deformation/rotation of the continuum under analytic transformations;
\item the interpretation of resonant states as (generalized) eigenstates in a spectral framework.
\end{itemize}
Complex scaling supplies the spectral setting; exact WKB supplies analytic continuation machinery and non-perturbative control.}

\section{Brief review of the exact WKB method}

\revone{Consider the $1$-dimensional stationary Schr\"odinger equation
\begin{equation}
\left[-\frac{\hbar^2}{2} \frac{d^2}{dx^2}+V(x)\right]\psi(x)=E\,\psi(x),
\end{equation}
and set
\begin{equation}
Q(x;E)=2[V(x)-E].
\end{equation}
A \textit{formal} WKB ansatz writes solutions as
\begin{equation}
\psi(x)=\exp\left(\int^x S(x';\hbar)\,dx'\right),
\end{equation}
where $S(x;\hbar)$ is expanded into a formal power series in $\hbar$,
\begin{equation}
S(x;\hbar)=\sum_{n=-1}^{\infty}\hbar^n S_n(x).
\end{equation}
Substituting into the Schr\"odinger equation yields the Riccati equation
\begin{equation}
S(x;\hbar)^2+\frac{dS(x;\hbar)}{dx}=\hbar^{-2}Q(x;E),
\end{equation}
which determines $S_n$ recursively. The leading term is
\begin{equation}
S_{-1}(x)=\pm\sqrt{Q(x;E)}.
\end{equation}
There exist two solutions depending on~$\pm$.
The points where $Q(x;E)=0$ are \textit{turning points}. The square-root $\sqrt{Q}$ is multivalued, so one works on the associated (typically hyperelliptic) Riemann surface. The well-known Proposition states that the WKB ansatz is given by
\begin{equation}
    \psi^{\pm}(x) = \frac{1}{\sqrt{S_{\mathrm{odd}}}} \exp\left(\pm\int^x S_{\mathrm{odd}}\, dx'\right) ,
\end{equation}
where $S_{\mathrm{odd}}\equiv \sum_{n\geq0} \hbar^{2n-1} S_{2n-1}$ with $S_{-1}=\sqrt{Q}$.}

\paragraph{\revone{Divergence and Borel resummation.}}
\revone{In general, the formal WKB series for $\psi_\pm$ is divergent (it is an asymptotic expansion).
The \textit{exact WKB method} upgrades the formal WKB series to actual analytic functions by
Borel resummation. One considers the formal series for $f(\hbar)=\sum_{n=0}^\infty a_n \hbar^n$
and forms its Borel transform in a variable $u$:
\begin{equation}
\mathcal{B}[f](u)=\sum_{n=0}^{\infty}\frac{a_n}{n!}u^n.
\end{equation}
This defines (by analytic continuation) a suitable function of $u$.
For a direction $\arg \hbar=\Theta\neq0$ such that the Laplace integral is well-defined,
the Borel sum is obtained by Laplace transform along the ray $u\in e^{i\Theta}\mathbb{R}_{\ge 0}$:
\begin{equation}
\mathcal{S}_{\Theta}[f](\hbar)
=\int_{0}^{e^{i\Theta}\infty} e^{-u/\hbar} \mathcal{B}[f](u)\,du.
\end{equation}
Applying this procedure consistently to the WKB exponent produces \textit{Borel-resummed WKB solutions}
$\Psi_{\pm}^{\Theta}(x;\hbar)$ that are genuine analytic solutions of the Schr\"odinger equation in appropriate domains. The dependence on $\Theta$ encodes Stokes phenomena as we will explain later.}

\paragraph{\revone{Stokes curves and Stokes graphs.}}
\revone{Turning points control where the Borel sum changes its analytic form.
Stokes curves are defined by the condition that
\begin{equation}
\im \int_{x_*}^{x} \sqrt{Q(x';E)}\,dx' = 0,
\end{equation}
where $x_*$ is a turning point. These curves emanate from turning points and form a \textit{Stokes graph},
which partitions the complex $x$-plane into regions where the resummed WKB solutions have uniform asymptotic behavior.}

\paragraph{\revone{Connection formulas and Stokes phenomena.}}
\revone{When $\hbar$ crosses a \textit{Stokes direction}, i.e., a direction for which the Borel transform develops pole singularities on the Laplace integration, the Borel sum jumps discontinuously:
\begin{equation}
    \Psi_{\pm}^{\Theta+0}-\Psi_\pm^{\Theta-0}\neq 0.
\end{equation}
This jump is governed by a Stokes automorphism and is expressed through \textit{connection formulas}
relating WKB solutions across Stokes curves. In practice,
$\Psi_\pm^{\Theta}$ and $\Psi_{\mp}^{\Theta}$ are dominant and subdominant, respectively, along a Stokes curve with $\re \revtwo{\hbar^{-1}}\int_{x_*}^x \sqrt{Q(x';\revtwo{E})} dx'\gtrless0$.
These formulas determine how a solution
that is subdominant (exponentially small) in one sector acquires an admixture of the dominant solution after crossing a Stokes curve.
This admixture is called the Stokes phenomenon accompanied with non-perturbative factors.}

\paragraph{\revone{Voros symbols and exact quantization.}}
\revone{A central exact WKB object is the \textit{Voros symbol}, defined (for a cycle $\gamma$ on the
Riemann surface of $\sqrt{Q}$) by the Borel-summed exponential of a period integral:
\begin{equation}
    \mathcal{V}_{\gamma}(\hbar)=
    \exp\left(\oint_{\gamma} \mathcal{S}_{\Theta}[S]\,dx\right),
\end{equation}
with $\Theta$ chosen in a non-singular direction.
\revtwo{The computation of the Voros symbol and the arguments here are straightforwardly applicable to general problems.}
Exact quantization conditions (for bound states or resonances) can often be written compactly in terms of
Voros symbols and their Stokes jumps, yielding non-perturbative information such as exponentially small
splittings and decay widths.}

\section{Transmission via exact WKB analysis}
\revone{For (pseudo-)bound states, exact WKB quantization of this kind has been well established. By contrast, in this paper we develop a general exact-WKB framework that extends to scattering states, i.e.\ the continuous spectrum.
In earlier works~\cite{Morikawa:2025grx,Morikawa:2025xjq}, the Voros symbols for the inverted Rosen--Morse potential were computed within the exact WKB framework, and were found to agree perfectly with the closed-form solution.\footnote{\revone{The techniques developed in Refs.~\cite{Morikawa:2025grx,Morikawa:2025xjq} can be extended in a straightforward manner, and they allow one to carry out Airy-type exact WKB computations also for general resonance potentials relevant to resonance physics.}} Here we present a methodology to analyze scattering problems for general scattering potentials by exploiting the geometric properties of the associated Stokes graphs.}

\subsection{\revone{Strategy to general scattering potentials}}
In scattering theory of a $1$-dimensional system, we first assume that an incoming wave, $e^{ikx}$, is made incident from $x\to-\infty$. Then, the asymptotic form of the corresponding wave function $\psi$ becomes $\psi(x\to-\infty) = e^{ikx} + R e^{-ikx}$ and $\psi(x\to\infty) = T e^{ikx}$, where $R$ and $T$ are the reflection and transmission coefficients. In general, for a $1$-dimensional problem, the transmission coefficient $T$ corresponds to the S-matrix and cross-section.

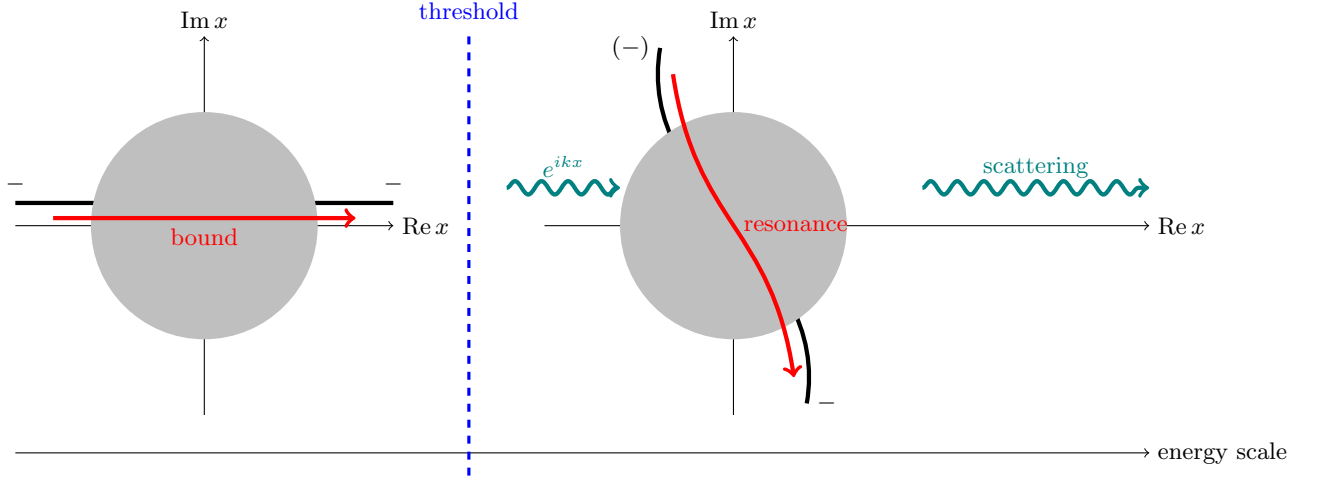
\begin{figure*}[t]
\centering
\begin{tikzpicture}
  \draw[->] (0,0) -- (15,0) node[right] {energy scale};

  \draw[->] (0,3) -- (5,3) node[right] {$\re x$};
  \draw[->] (2.5,0.5) -- (2.5,5.5) node[above] {$\im x$};

  \draw[ultra thick] (2.5,3.3) -- (0,3.3) node[above] {$-$};
  \draw[ultra thick] (2.5,3.3) -- (5,3.3) node[above] {$-$};

  \fill[gray!50] (2.5,3) circle (1.5);

  \draw[ultra thick,red,->] (0.5,3.1) -- (4.5,3.1) node[midway,below] {bound};

  \draw[very thick,dashed,blue] (6,-0.3) -- (6,5.6) node[above] {threshold};

  \draw[->] (7,3) -- (15,3) node[right] {$\re x$};
  \draw[->] (9.5,0.5) -- (9.5,5.5) node[above] {$\im x$};

  \draw[ultra thick] (9.5+1,3) arc(270:170:2) node[left] {$(-)$};
  \draw[ultra thick] (9.5-1,3) arc(90:-10:2) node[right] {$-$};

  \fill[gray!50] (9.5,3) circle(1.5);

  \draw[ultra thick,red,->] (9.5-0.8,5) .. controls (9.5-0.5,3) and (9.5+0.5,3) .. (9.5+0.8,1) node[midway,right] {resonance};

  \draw[ultra thick,teal,snake it,->] (6.5,3.5) -- (8,3.5) node[midway,above] {$e^{ikx}$};
  \draw[ultra thick,teal,snake it,->] (12,3.5) -- (15,3.5) node[midway,above] {scattering};
\end{tikzpicture}
  \caption{Schematic diagram of the energy scale for Stokes graphs. For energies below the threshold ($E \leq 0$), bound states are present, with the Stokes curve with index $-$ extending towards $\pm \infty$ along the red path, ensuring the normalizability of the discrete spectrum. In the gray region, the appropriate connection formula applies, and branch cuts may be needed. Beyond the threshold, where the potential becomes flat at large $\re x$, the Stokes curve extends in the $\im x$ direction. With the correct sheet choice and the Stokes curve with index $-$, normalizability along the red path is guaranteed, revealing resonances with a discrete but complex spectrum. The choice of path corresponds to the Siegert boundary condition, where incoming waves are prohibited. Assuming an incoming wave $e^{ikx}$ from $\re x \to -\infty$, scattering states (transmitted and reflected waves) appear, corresponding to the continuum spectrum.}
  \label{fig:stokes_dist}
\end{figure*}

In the exact WKB analysis, we also impose boundary conditions where the incident wave does not vanish to give continuum spectrum, while bound/resonant states are normalizable on an appropriate choice of analytically continuing path. Figure~\ref{fig:stokes_dist} shows that (i) for bound energies, say $E<0$, there exist the Stokes curves, where $\psi^{-}\propto e^{-\int S_{\mathrm{odd}}dx}$ is dominant, at $x\to\pm\infty$; one may observe a nontrivial Stokes structure, branch cut, and monodromy in the middle region, which gives rise to quantized energy spectrum based on the normalizability of wave functions.
(ii) Above an energy threshold ($\re E>0$), the flat potential $V(x)\sim 0$ at $x\to\pm\infty$ indicates that there is no Stokes curves with index~$\pm$ at $|\re x|\to\infty$, but we see Stokes curves extending to $|\im x|\to\infty$ and a nontrivial structure in between; Refs.~\cite{Morikawa:2025grx,Morikawa:2025xjq} addressed the quantized complex energies of resonance from the normalizability along these Stokes curves. (iii) Now, we expect that wave functions with monodromy applied to $e^{ikx}$ will become the scattered wave.

The strategy to compute this scattering problem is as follows:
\begin{enumerate}
    \item Specifically, impose $\psi(x\to\infty) = 1 \times e^{ikx}$.
    \item This can be implemented as the boundary condition which is $\psi^{-}$-dominant as $\im x\to-\infty$ in $\re x>0$ (see Fig.~\ref{fig:stokes_dist} and discuss it as the Siegert boundary condition later).
    \item By working on exact WKB framework, we obtain $\psi^{\pm}(x\to-\infty)$.
    \item The coefficient for the incident wave $e^{ikx}$ is $1/T$, and one of the reflected wave $e^{-ikx}$ is $R/T$.
\end{enumerate}
From this, we can observe the S-matrix, phase shift, and the Breit--Wigner formula straightforwardly.

\subsection{\revone{Transmission coefficient in Rosen--Morse potential}}
For instance, let us focus on the Rosen--Morse potential,
$\frac{U_{0}}{\cosh^2 \beta x}$ with $U_0>0$.
The general solution of its Schr\"{o}dinger equation is described by using the two linearly independent solutions as
\begin{align}
    &\psi(x) \notag \\
    &= 
    (1-\xi^2)^{-\frac{ik}{2\beta}}
    \notag \\
    &\quad\times\Biggl[
    C_1 F\left( -\frac{ik}{\beta}-s, -\frac{ik}{\beta}+s+1, -\frac{ik}{\beta}+1, \frac{1-\xi}{2} \right)
    \notag \\
    &\quad\qquad
    + C_2\left(\frac{1-\xi}{2}\right)^{\frac{ik}{\beta}}
    F\left( s+1, -s, \frac{ik}{\beta}+1, \frac{1-\xi}{2} \right)
    \Biggr],
 \label{eq:rm_sol}
\end{align}
where $F$ is the Gaussian hypergeometric function,
$\xi=\tanh{\beta x}$, $k=\frac{\sqrt{2mE}}{\hbar}$, and $s=\frac{1}{2}\left( -1+\sqrt{1-\frac{8mU_{0}}{\beta^2 \hbar^2}} \right)$.
The first and second terms correspond to the outgoing and incoming wave, respectively, in the asymptotic region $x\to\infty$.

\subsubsection{Airy-type local model (simple turning point).}
\revone{Let $Q(x;E)=V(x)-E$ and suppose $x_t$ is a \textit{simple} turning point, i.e.
\begin{equation}
Q(x_t;E)=0,\qquad Q'(x_t;E)\neq 0.
\end{equation}
Expanding $Q$ near $x_t$ gives
\begin{equation}
Q(x;E)=Q'(x_t;E)(x-x_t)+O\!\left((x-x_t)^2\right).
\end{equation}
Keeping only the linear term and rescaling the coordinate reduces the Schr\"odinger equation locally to an Airy
equation. This is the origin of the standard Airy-type connection formulas.
Consequently, the Airy reduction is reliable only when the region relevant for matching is contained in a domain
where the higher-order terms in $Q$ are negligible.}

Now, following the above strategy, we take $C_1=4^{\frac{ik}{2\beta}}$ and $C_2=0$ to impose the normalized transmitted wave; we see $\psi(x)\to e^{ikx}$ as $x\to\infty$. As a consequence of the Airy-type exact WKB analysis, we obtain
\begin{align}
  \frac{1}{T_{\text{Airy}}} &= \frac{1}{\sqrt{A}} - \sqrt{A},\\ 
   A &= \left[
    \frac{F\left( -\frac{ik}{\beta}-s, -\frac{ik}{\beta}+s+1, -\frac{ik}{\beta}+1, \frac{1-\Tilde{\xi}}{2} \right)}
    {F\left( -\frac{ik}{\beta}-s, -\frac{ik}{\beta}+s+1, -\frac{ik}{\beta}+1, \frac{1+\Tilde{\xi}}{2} \right)}
    \right]^2 .
\end{align}
Here $\Tilde{\xi}=\tanh(\cosh^{-1}\sqrt{U_0/E})$,
\revtwo{and the corresponding Voros symbol~$\mathcal{V}_\gamma$ (called $A$-cycle), where a non-perturbative cycle~$\gamma$ encircles two turning points $a_1$ and $a_2$ and is defined as~$A=e^{\oint_\gamma dx\, S_{\mathrm{odd}}}$,} is computed in Refs.~\cite{Morikawa:2025grx,Morikawa:2025xjq}.
\revthr{The $A$-cycle has originally been used in the quantization condition for resonant states. Here we use it to express the Airy-type approximation to the transmission coefficient $T_{\text{Airy}}$ and to specify the corresponding normalized scattering solution (with $C_1=4^{ik/(2\beta)}$ and $C_2=0$).}
Note that $T$ is divergent if $E$ comes across resonances; so a naive analytical continuation of the wave function is ill-defined in some sense, such as completeness. A Hilbert space can be constructed by energy eigenstates with $\im E=0$, where the $*$-operation is well-defined and the Hamiltonian is self-adjoint; now, resonance is not a ``state'' but just a physical pole singularity. To include resonant states in our Hilbert space, we need to introduce CSM and show the ABC theorem.

\subsubsection{Comparison with exact solution}
\begin{figure}
    \centering
    \includegraphics[width=0.9\linewidth]{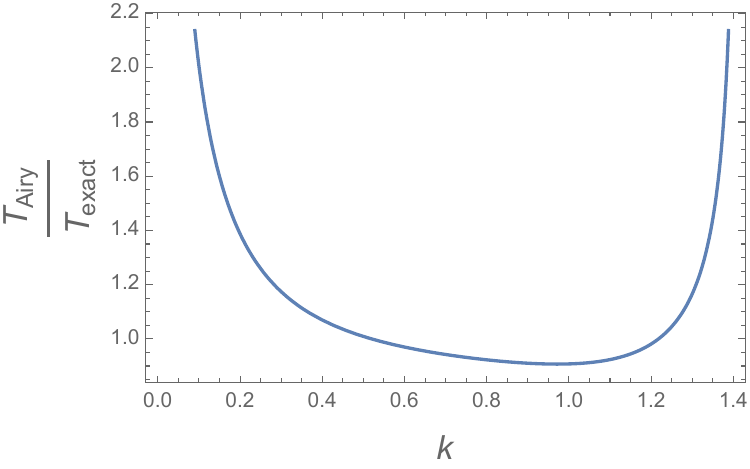}
    \caption{Ratio of the transmission coefficients, $|T_{\text{Airy}}|/|T_{\text{exact}}|$. Every parameter excluding~$E$ is set to unity. $k$ and $\Tilde\xi$ are written by~$E$, and the $k$-dependence of the ratio is shown. $k=0$ and $\sqrt{2}$ correspond to $E=0$ and $E=U_0=1$, respectively.}
    \label{fig:trans-coeff}
\end{figure}

From the asymptotic expansion of the exact solution~\eqref{eq:rm_sol} near $x\to\pm\infty$, on the other hand, it is well-known~\cite{Landau:1991wop} that the transmission coefficient is rigorously given by
\begin{align}
    T_{\text{exact}}=
    \frac{\Gamma\left(-\frac{ik}{\beta}-s\right) \Gamma\left(-\frac{ik}{\beta}+s+1\right)}{\Gamma\left(1-\frac{ik}{\beta}\right) \Gamma\left(-\frac{ik}{\beta}\right)} .
\end{align}
The approximation by the Airy function is hindered by non-linearity around stationary points, and so works well at enough large~$E>0$ but at enough small $E<U_0$.\footnote{In fact, a preliminary numerical simulation observes a slight discrepancy in the frequency of the quasi-normal mode in a black hole, while the resonant peak is completely reproduced~\cite{Miyachi:25}.} See Fig.~\ref{fig:trans-coeff}. One may use the degenerate Weber to tackle $E\sim0$ or $U_0$. In any case, it is remarkable that bound/resonant energy eigenvalues are sufficiently far from the stationary, and given by the exact WKB analysis rigorously.

\revone{If the linear approximation of $Q(x;E)$ is poor (for instance, due to sizable curvature or because the effective
barrier top is sampled), it is natural to replace the Airy normal form by a quadratic (parabolic) normal form.
A prototypical choice is a Weber-type (parabolic-cylinder) equation, which captures the local behavior of
\begin{equation}
Q(x;E)\simeq a(x-x_0)^2+b
\end{equation}
and yields uniform connection data for ``parabolic'' turning-point configurations.
In such a formulation, the local model remains accurate even when the linearized (Airy) neighborhood is too small,
and one expects improved agreement with the exact solution in parameter regimes where Fig.~\ref{fig:trans-coeff} shows strong deviations.}

\subsubsection{Remark on cycles and matching data.}
\revone{Since different local normal forms (Airy, Weber, etc.) correspond to different choices of local coordinates and
branch structures, the associated natural cycles (e.g.\ an ``$A$-cycle'' used to define phase/connection data)
need not coincide and may differ substantially except for poles.
Therefore, the discrepancy in Fig.~\ref{fig:trans-coeff} should be interpreted as indicating that the Airy-type local model is not
optimal for the chosen parameters, and that a Weber-type (or other higher-order) local treatment may be required.}

\revone{This improvement is conceptually independent of whether one employs a Zinn-Justin-type semiclassical expansion
or a Borel-resummed (exact WKB) framework; it concerns the choice of an appropriate local normal form for $Q(x;E)$. This remains an important open problem.}

\section{ABC theorem in exact WKB framework}
\subsection{\revtwo{Derivation of ABC theorem}}
Let us show the most crucial theorem in CSM, the ABC theorem, from the perspective of the exact WKB analysis. At first, it is shown~\cite{Morikawa:2025grx} that the resonant wave function is square-integrable, i.e., $L_2(\Tilde{\mathbb{R}})$, where $\Tilde{\mathbb{R}}$ is defined on the red curved path in Fig.~\ref{fig:stokes_dist}. CSM via exact WKB can also provide the wave functions in $L_2(\mathbb{R})\simeq L_2(\Tilde{\mathbb{R}})$, where $\mathbb{R}=(-\infty,\infty)$~\cite{Morikawa:2025xjq}.

Second, from the discussion in Ref.~\cite{Morikawa:2025xjq}, it is straightforward to show the $\theta$-independence of bound/resonant energy spectrum because Stokes curves, any (non-)perturbative cycle and turning points are rotated simultaneously by the complex scaling, and hence the normalizability condition on Stokes curves is invariant \revone{and provides the same exact-WKB quantization}.
\revone{For instance, the non-perturbative cycle $A$ is now given by
\begin{align}
    A &= \exp\oint_{\revtwo{\gamma e^{-i\theta}}} S_{\mathrm{odd}}^\theta
    = \left[\frac{\psi_{\mathrm{CSM}}(a e^{-i\theta})}{\psi_{\mathrm{CSM}}(-a e^{-i\theta})}\right]^2\notag\\
    &= \left[\frac{\psi(a)}{\psi(-a)}\right]^2
    = \exp\oint_{\revtwo{\gamma}} S_{\mathrm{odd}} ,
\end{align}
where $\psi_{\mathrm{CSM}}(x' e^{-i\theta}) = \psi(x')$, \revtwo{$\gamma e^{-i\theta}$} denotes the \revtwo{cycle between two turning points} after complex rotation, and the original turning point \revtwo{is given} with $\theta=0$.}
This fact is nothing but the invariance of the energies of the bound and resonant states.

Third, the above exact WKB strategy for scattering problem is also valid in CSM. Namely, when $x\to x e^{i\theta}$, we should take $k\to k e^{-i\theta}$ to maintain the transmission wave~$e^{ikx}$ as the boundary condition.
As mentioned in the case of resonance, the dominance of~$\psi^{-}$ at $|\im x|\to\infty$ indicates a physical situation such that any incoming wave vanishes.
Hence, this boundary condition requires that $\psi^-$ is dominant as $\im x \to - \infty$ in $\re x >0$, while eliminating $\psi^+$; this can be interpreted as the physical meaning of the Siegert boundary condition.\footnote{Here we are concerned solely with the physical interpretation of the Siegert boundary condition. Normally, this condition corresponds to eliminating $\psi^+$ for $\re x <0$. However, this is equivalent to requiring that $\psi^-$ be dominant, as in the case of $\re x >0$.}
Now, since $k=\frac{\sqrt{2m E}}{\hbar}$, the continuum spectrum $E_{\mathrm{cont}}$
is changed as $E_{\mathrm{cont}}\to E_{\mathrm{cont}} e^{-2i\theta}$.


\subsection{\revtwo{Equivalence with Feshbach resonance}}

\revone{We here make a remark on an equivalence with the Feshbach picture.
The gray circle in Fig.~\ref{fig:stokes_dist} should be understood as a projection of the full scattering problem into an ``interior'' (gray-filled) sector and an ``exterior'' sector.
More concretely, let $P$ denote the degrees of freedom supported in the interior region that contains the nontrivial (resonant) Stokes graph.
On the other hand, the dynamics in the exterior region, $Q=1-P$, is asymptotically free, and so the analytic structure is essentially trivial.
In the Feshbach projection formalism, eliminating the $Q$-sector yields an energy-dependent effective Hamiltonian operator acting on the interior sector,
\begin{align}
  H_{\mathrm{eff}}(E)=PHP + PHQ\frac{1}{E-QHQ}QHP .
\end{align}
It has been known~\cite{Feshbach1958UnifiedI,Feshbach1962UnifiedII,Rotter:2007zng} that non-Hermiticity encodes the open boundary coupling to the environment, and complex poles of the effective resolvent of~$H_{\mathrm{eff}}$ reproduce resonances.}

\revone{In the present exact WKB, the $Q$-elimination affects an effective boundary condition imposed at the edge of~$P$.
Choosing an outgoing (Siegert-type) condition excludes incoming components from the exterior. It thus implements precisely the same ``radiation condition'' that is built into $H_{\mathrm{eff}}(E)$ through the resolvent $(E-QHQ)^{-1}$ \cite{Hatano:2013nyz}.
From the exact-WKB quantization condition (the normalizability at $|\im x|\to\infty$), now the connection data across the interior Stokes graph matches the outgoing boundary condition at the edge, that is, a vanishing Jost function or a pole of the scattering amplitude.
Accordingly, the exact-WKB resonant state can be viewed as a bound state along $\im x$, and as the counterpart of the Feshbach-pole condition (openness along $\re x$ and a pole of $(E-H_{\mathrm{eff}}(E))^{-1}$).

\revtwo{%
We would like to clarify what we mean by the $P/Q$ ``cut'' and its relation to the Feshbach picture.
To avoid possible confusion, we spell out the viewpoint and the logical order of definitions.
\begin{itemize}
  \item \textbf{Guiding idea (assumption).}
  We adopt the standard standpoint that the \textit{full} system ($P+Q$) is governed by a Hermitian operator,
  while a \textit{subsystem} becomes effectively non-Hermitian once the environmental/continuum sector is eliminated,
  i.e.\ openness is encoded in an energy-dependent effective description.
  \item \textbf{\textit{A posteriori} Hilbert space in exact WKB.}
  In the exact WKB approach, the primary data are analytic solutions on the complexified coordinate plane and their Stokes geometry.
  A Hilbert-space structure (and hence ``normalizability'') is typically not assumed from the outset, but rather introduced \textit{a posteriori} through a complex deformation (via the ABC/complex-scaling viewpoint) and the resulting decay/normalizability requirement.
  \item \textbf{Definition of the $Q$ sector.}
  We define $Q$ as the exterior region which is expected, from the Stokes-graph viewpoint, to carry no additional nontrivial connection data (in this sense its analytic structure is essentially trivial), and set $P=1-Q$.
  Equivalently, $P$ is the region containing the nontrivial (resonant) Stokes graph.
  This $P/Q$ decomposition should be regarded as a \textit{partition of the complex $x$-plane} rather than an \textit{a priori} Hilbert-space decomposition.
  \item \textbf{Remark (not yet exact WKB).}
  In the usual Feshbach projection formalism, after eliminating $Q$ resonances correspond to complex poles of the effective resolvent $(E-H_{\mathrm{eff}}(E))^{-1}$.
  \item \textbf{Proposition.}
  Imposing an outgoing (Siegert-type) condition at the $P/Q$ boundary is equivalent to describing an open system with \textit{no influx from $Q$ into $P$} and \textit{only outflux from $P$ into $Q$}.
  \item \textbf{Remark.}
  It is known that the resonance spectrum selected by the Siegert boundary condition coincides with that obtained from the radiative/open effective Hamiltonian $H_{\mathrm{eff}}(E)$.
  \item \textbf{Statement (exact WKB language).}
  The ``matching'' of exact WKB solutions in $P$ and in $Q$ across the boundary means the following:
  an exact-WKB solution in $P$ that is normalizable as $|\im x|\to\infty$ can be smoothly connected at the $P/Q$ boundary to the (almost analytically trivial) outgoing solution in $Q$ along the physical direction $\re x$.
  Geometrically, this is precisely the condition that the interior Stokes connection data be compatible with the exterior outgoing condition, i.e., the exact-WKB characterization of resonances.
\end{itemize}}

\revtwo{In particular, the exact WKB approach allows us to characterize all these states in terms of global analytic structures of the wave function, such as Stokes graphs and connection formulas. From this viewpoint, the Siegert boundary condition can be reinterpreted as the Stokes curve emanating from $|\im x|\to\infty$. In the Feshbach formalism it can be understood as a consequence of analytic continuation in the complex plane (no Stokes curve along $\re x$ axis), rather than as an external imposition.}

\revtwo{The conflict between the directions of Stokes curves connected with infinity and the real axis implies that the notion of an ``outgoing wave'' along the real coordinate becomes nontrivial and must be defined via analytic continuation along appropriate Stokes sectors. This is closely related to the emergence of resonant (Siegert) states and their interpretation within the Feshbach formalism, where one effectively projects onto subspaces associated with different asymptotic behaviors.}

In this sense, Fig.~\ref{fig:stokes_dist} provides a geometric realization of the resonance mechanism in the Feshbach formalism: interior (quasi-)bound structures defined by the resonant Stokes graph acquire decay widths through their coupling to the exterior continuum.}

\section{Continuum and resonant spectra in Hilbert space}
\revone{In the previous section, our aim is to \textit{physically interpret} the Aguilar--Balslev--Combes (ABC) theorem using the exact WKB framework,
then establishing a completeness statement plays a central role for three reasons.
\begin{itemize}
\item \textbf{Correct extraction of resonance.}
Exact WKB naturally produces local solutions and their connection data (via Stokes graphs).
However, the ABC picture is a \textit{global spectral statement}.
A completeness relation makes this precise by stating that the identity (or the resolvent matrix elements) can be represented by
(rotated continuum) + (sum over resonant poles) + (bound states).
The ``extraction'' of resonance contributions is an actual reorganization of the spectral representation.
\item \textbf{Consistent Hilbert-space setting.}
\revtwo{Resonant} states are not $L^2$ on the real axis, so they do not enter the standard Hilbert-space completeness for the self-adjoint $H$.
In contrast, after complex scaling they may become $L^2$ eigenstates of the non-self-adjoint $H_\theta$.
A completeness statement clarifies \textit{in which sense} resonant states are part of the spectral decomposition:
typically via a biorthogonal resolution of the identity for $H_\theta$ or, equivalently, via contour deformation and residues for resolvent matrix elements.
Without this, the claim ``resonances appear in the spectrum'' risks being interpreted incorrectly as a literal
Hilbert-space basis statement for the original $H$.
\item \textbf{Physical observables.}
Exact WKB yields scattering data and resonance conditions through Stokes phenomena (connection matrices, Voros symbols).
To relate these quantities to physical observables (e.g.\ cross sections, spectral functions, time evolution),
one needs a representation of the Green function or evolution kernel in which the contributions of continuum and resonance sectors are clearly separated and controlled.
Completeness provides exactly that bridge: The residues and pole locations are the resonance data that exact WKB computes.
This makes precise why the Stokes-graph geometry (WKB side) reproduces the ABC spectral separation (CSM side).
\end{itemize}
In an exact-WKB-based interpretation of ABC, ``showing completeness'' is meaningful because it is the statement
that the WKB-derived resonance poles and connection coefficients are not just diagnostic features, but
\textit{enter the actual spectral representation} (resolvent or identity) together with the rotated continuum.
It is the step that upgrades a physical picture (``continuum vs.\ resonances'') into a mathematically and physically controlled decomposition.
}

\revtwo{It should be emphasized that, according to the picture we have presented, all wave functions---bound, scattering, and resonant---can be derived within the framework of the exact WKB method. Therefore, the notion of completeness referred to here is not that of the conventional scattering theory of the Schr\"odinger equation; rather, completeness must be formulated entirely in terms of wave functions derived via the exact WKB method. Establishing such a completeness is essential for completing the overall structure of exact WKB scattering theory.}

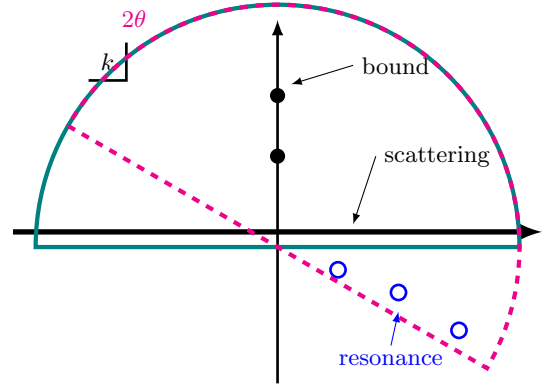
\begin{figure}[t]
    \centering
  \begin{tikzpicture}[x=1mm,y=1mm,>=latex]
   \draw[->,line width=2pt] (-35,0) -- (35,0);
   \draw[->,line width=1pt] (0,-20) -- (0,28);
   \draw[line width=1pt] (-25,20) -- (-20,20) -- (-20,25);
   \draw[line width=1pt] (-22.5,22.5) node {$k$};
   \draw[line width=1pt,blue] (8,-5) circle (1);
   \draw[line width=1pt,blue] (16,-8) circle (1);
   \draw[line width=1pt,blue] (24,-13) circle (1);
   \fill (0,10) circle (1);
   \fill (0,18) circle (1);
   \draw[<-] (1.8,19.5) -- (10,22) node[right] {bound};
   \draw[<-] (10,1) -- (13,10) node[right] {scattering};
   \draw[<-,blue] (16,-10.5) -- (15,-15) node[below] {resonance};
   \draw[teal,ultra thick] (-32,-2) -- (32,-2) arc(0:180:32) -- cycle;
   \draw[magenta,dashed,ultra thick,rotate around={-30:(0,-2)}] (-32,-2) -- (32,-2) arc(0:180:32) -- cycle node[above left] {$2\theta$};
  \end{tikzpicture}
    \caption{Integration contour in the complex $k$-plane of the Green function. The green contour indicates the original integration to prove the completeness~\eqref{eq:complete_no-res}. We find that in $\im k<0$ there exist resonant poles additionally. An integration along the magenta contour rotated by $2\theta$ via CSM concludes these residues as a physical state depending on~$\theta$.}
    \label{fig:complete_contour}
\end{figure}

We introduce the standard Green function satisfying
\begin{align}
    H G(k;x,x') = -\delta(x-x') ,
\end{align}
with a Hamiltonian operator~$H$.
One may write it as the resolvent,
\begin{align}
    G(k;x,x') = \int d\rho(E') \frac{\psi(x)\psi(x')^*}{E-E'} ,
    \label{eq:resolvent}
\end{align}
where \revone{$\psi(x)$ is a generic wave function, which collectively denotes the bound and scattering states, before normalization and} we should be careful of the branch cut to obtain the outgoing/incoming Green function, and have introduced the spectral function
\begin{align}
    \frac{d\rho(E)}{dE} \propto
    \begin{cases}
        \sum_{n} \frac{1}{(\mathcal{N}_n^B)^2} \delta(E-E_n^B) & E\leq0,\\
        \frac{1}{|\mathcal{F}_k|^2} \revone{\times} (\mathrm{const.}) & E>0 ,
    \end{cases}
\end{align}
the bound energies~$E_n^B$ ($n\in\mathbb{Z}_+$), the normalization factor of the bound states \revone{$\mathcal{N}^B=(\int|\psi(x)|^2dx)^{1/2}$ for bound states~$\psi(x)$}, and the coefficient of waves in the continuum spectrum $\mathcal{F}_k$, \revone{that is, the Jost functions}.
From the self-consistency condition of an integral of the Green function given in Ref.~\cite{Newton:1960}, where we take into account the integration on the upper half $k$-plane surrounded by the solid line in Fig.~\ref{fig:complete_contour}, the completeness of the wave functions is proved as
\begin{align}
    \sum_n\psi_n^B(x)\psi_n^B(x')^* + \int dk\, \psi(k,x)\psi(k,x')^* = \delta(x-x') .
    \label{eq:complete_no-res}
\end{align}
Here, $\psi^B(x)$ implies a \revone{normalized} bound state, and $\psi(k,x)$ is a \revone{normalized} scattering one. If there exist resonances, the resolvent has pole singularities after an analytic continuation of~$k$ to $\im k<0$.

In what follows, we consider the complete set of eigenstates in CSM QM. First of all, we notice that the Green function~\eqref{eq:resolvent} is quite subtle at a resonant pole since its residue, i.e., the spectral function, diverges due to $\mathcal{F}_k=0$. \revone{In a mathematical context, the spectral function \revtwo{for the continuum spectrum} is defined in a band around~$\im E=0$ such that $\mathcal{F}_k\neq0$. Nevertheless, under CSM, resonant states are also elements of the corresponding spectral representation.} (Also, there are subtleties for the definition of the $*$-operation in Eq.~\eqref{eq:resolvent}.\footnote{The original (unscaled) Hamiltonian is self-adjoint on an appropriate domain, ensuring a real spectrum and a well-defined Banach-space-valued resolvent. CSM and the ABC theorem guarantee that, after analytic dilation, the rotated Hamiltonian remains a closed, sectorial operator with resonances and a well-behaved resolvent~\cite{Davies:07}. Although self-adjointness is lost after complex scaling, the operator remains diagonalizable and retains a modified form of elliptic regularity~\cite{Gelfand:1964}.
In genuinely non-Hermitian systems where such analytic control is absent, one must explicitly assume or prove diagonalizability and the existence of Riesz projectors~\cite{Trefethen1999,Davies:2002}, since no elliptic regularity is automatically guaranteed.})

\revone{Note that $\psi(x)^\theta=U_\theta\psi(x)$ for a general wave function before normalization. Let $\psi^B(x)^\theta$ denote a \revone{normalized} complex-scaled bound state, $\psi(k,x)^\theta$ is a \revone{normalized} complex-scaled scattering one, and $\psi^R(x)^\theta$ imply a \revone{normalized} resonant state.}
We shall define the CSM Green function
\begin{align}
    G_\theta(k;x,x') = \int d\rho_\theta(E') \frac{\psi(x)^\theta\psi(x')^\theta}{E-E'} ,
    \label{eq:resolvent_csm}
\end{align}
and the CSM spectral function
\begin{align}
    &\frac{d\rho_\theta(E)}{dE} \notag\\ &\propto
    \begin{cases}
        \sum_{n} \frac{1}{(\mathcal{N}_n^B)^2} \delta(E-E_n^B) & E\leq0\\
            & \revtwo{\text{(bound)}},\\
        \frac{1}{|\mathcal{F}_k|^2} \revthr{f_\theta(E)} & \revone{\arg E=-2\theta} \\
            & \revtwo{\text{(rotated conti.)}},\\
        \sum_{m} \frac{1}{(\mathcal{N}_m^R)^2} \delta(E-E_m^R) [1-\revthr{f_\theta(E)}] & \im E<0\\
            & \revtwo{\text{(resonant)}}, 
    \end{cases}
\end{align}
\revthr{where $\mathcal{N}^R$ is the normalization factor of resonant states in exact WKB analysis, and $E_m^R$ denotes the resonance energies. For each resonance energy $E_m^R$, we define the critical scaling angle by
\begin{align}
    \theta_R^m
    =
    \frac{1}{2}
    \arctan\left(
        -\frac{\im E_m^R}{\re E_m^R}
    \right) .
\end{align}
Here $\theta_R^m$ is the value of $\theta$ at which the rotated continuum line $\arg E=-2\theta$ passes through the resonance energy~$E_m^R$.
For a fixed scaling angle $\theta$, the resonant contribution is summed over all $m$ such that
\begin{align}
    \theta_R^m < \theta .
\end{align}
Equivalently, if the critical angles are ordered as
\begin{align}
    \theta_1^R < \theta_2^R < \cdots ,
\end{align}
then the resonant states included in the spectral decomposition are precisely those with $m=1$, \dots, $M(\theta)$, where $M(\theta)$ is determined by
\begin{align}
    \theta_{M(\theta)}^R < \theta < \theta_{M(\theta)+1}^R .
\end{align}
The borderline case $\theta=\theta_m^R$ requires separate care, since the resonance energy lies on the rotated continuum line.
$f_\theta(E)$ denotes a $\theta$-dependent support function on the spectrum. It equals $1$ on the rotated continuum part that remains in the continuum representation, and vanishes at those resonance energies $E_m^R$ for which $\theta_R^m<\theta$, i.e. for resonances that have already been isolated from the continuum and are counted separately in the discrete resonant sum.}

The integration is carried out along the dashed line in Fig.~\ref{fig:complete_contour}. In this case, the ill-definedness \revone{(divergences due to $\mathcal{F}_k=0$)} of the spectral function is remedied \revone{because of the support function~\revthr{$f_\theta(E)$}}\footnote{\revone{We assume that $f_\theta(E)$ is a smooth cutoff function.
If no support function, the case that $\theta=\theta_R$ becomes an exceptional point; not only the energy eigenvalues but also the energy eigenvectors are degenerate (or defective/Jordan-blocking). Hence, the calculation requires the utmost caution. In the more recent work~\cite{Morikawa:2025inx}, we have developed non-Hermitian physics of exceptional point for resonant and continuum spectra under CSM.}}, and near $\theta\revone{\gtrsim}\theta_R$ the scattering $\psi(k,x)^\theta$ is \revone{reorganized} by the corresponding resonant state $\psi^R(x)^\theta$ \revone{in the sense of spectral decomposition}, both of which have an identical energy eigenvalue. Again, one may say that the divergence of the scattering state originates from the divergent coefficient of the incoming wave, and its appropriately normalized state is resonant with the Siegert boundary condition.
\revthr{When $\theta$ crosses $\theta_R^m$, the rotated continuum itself does not disappear; rather, only the corresponding resonance energy $E_m^R$ is extracted from the continuum contribution and is thereafter counted in the discrete resonant part.}
\revthr{Thus, for $\theta>\theta_R^m$, the corresponding resonance is incorporated into the discrete resonant part of the spectral decomposition, while the rotated continuum contribution is modified accordingly.}

Finally, \revone{after defining our CSM spectral function,
the completely identical approach to the self-consistency condition of integration of the Green function in Ref.~\cite{Newton:1960} is valid; similar calculations leads to our central result}
\begin{align}
    &\sum_n\psi_n^B(x)^\theta\psi_n^B(x')^\theta + \int_{\revthr{\arg E=-2\theta}} dk\, \psi(k,x)^\theta\psi(k,x')^\theta \notag\\
    &\qquad
    + \sum_{m\revthr{\,:\,\theta_R^m<\theta}}\psi_m^R(x)^\theta\psi_m^R(x')^\theta
    = \delta(x-x') .
    \label{eq:complete_res}
\end{align}
\revtwo{Here, all bound, scattering, and resonant wave functions are exact WKB-derived.}
\revthr{The sum over $m$ is taken over all resonances satisfying $\theta_R^m<\theta$, namely $m=1$, \dots, $M(\theta)$ in the above ordering.}
Note that the summation over~$m$ depends on~$\theta$, that is, whether the integral contour rotated by~$2\theta$ includes the poles or not (see Fig.~\ref{fig:complete_contour}).

\section{Conclusion}
In this work, we have applied the complex scaling method (CSM) and exact WKB analysis to rigorously describe both continuum and resonant spectra in quantum systems. By reinterpreting the Aguilar--Balslev--Combes (ABC) theorem in the context of exact WKB, we have demonstrated how resonant states, traditionally seen as poles of the S-matrix, can be effectively incorporated into the Hilbert space through the introduction of a modified boundary condition. This work provides a clear framework for analyzing scattering problems, where the exact WKB analysis allows for precise computation of transmission coefficients and scattering cross-sections.

We also explored the physical significance of the Siegert boundary condition, which restricts incoming waves for resonant states, and how this boundary condition plays a crucial role in defining the normalizability of resonant wavefunctions. Our approach bridges the gap between phenomenological descriptions and rigorous mathematical formulations, allowing for the inclusion of resonant states in a self-consistent manner.

Finally, we have shown that the continuum and resonant spectra can be treated within the same formalism, with the exact WKB analysis serving as a powerful framework for understanding their interrelations. The completeness of the set of eigenstates, including bound, resonant, and scattering states, is established, and the theory is extended to handle the physical singularities. Our results provide new insights into scattering theory and resonance phenomena, with potential applications in various areas of quantum physics.

\begin{acknowledgments}
We are grateful to Masazumi Honda and Yuto Moriwaki.
We would like to thank the officers in Nishijin Plaza at Kyushu University, and the restaurant Ajisawa for their hospitality.
This work was partially supported by Japan Society for the Promotion of Science (JSPS)
Grant-in-Aid for Scientific Research Grant Numbers
JP25K17402 (O.M.), JP21H04975 (S.O.), and 25H01267 (S.O.).
O.M.\ acknowledges the RIKEN Special Postdoctoral Researcher Program
and RIKEN FY2025 Incentive Research Projects.
\end{acknowledgments}

\bibliographystyle{apsrev4-1}
\bibliography{ref,ref_ewkb,ref_res}
\end{document}